# Comprehensive investigation on the correlation of growth, structural and optical properties of GaN nanowires grown on Si(111) substrates by plasma assisted molecular beam epitaxy technique

Ritam Sarkar[a], R. Fandan[a], Krista R. Khiangte[b], S. Chouksey[a], A. M. Josheph[a], S. Das[a], S. Ganguly[a], D. Saha[a] and Apurba Laha[a, †]

[a]Department of Electrical Engineering, Indian Institute of Technology Bombay, Powai, Mumbai, India

[b]Department of Physics, Indian Institute of Technology Bombay, Powai, Mumbai, India.

† Corresponding author: A. Laha, Email: laha@ee.iitb.ac.in.

**Abstract**

The present study elucidates the correlation between the structural and optical properties of GaN nanowires grown on Si(111) substrate by plasma assisted molecular beam epitaxy (PA-MBE) technique under various growth conditions. GaN NWs exhibiting different shapes, sizes and distribution were grown at various substrate temperatures with same Ga-N (III-V) ratio of 0.4. We observe that sample grown at lower substrate temperature (~700°C) results 2-dimensional island like structure with almost very little (~30%) circularity while increasing substrate temperature (>770°C) leads to growth of individual GaN NW (>80% circularity) with excellent structural properties. The temperature dependent photoluminescence measurement together with analysis of RAMAN active modes provides legitimate evidences of strong correlation between the structure and optical properties GaN nanowires grown on Si substrates.

**Introduction**

GaN nanowires (NWs) exhibit extraordinary potential in recent days optoelectronic devices such as solid state lighting (light emitting diode, laser), energy harvesting devices (photovoltaics) and high power, high frequency devices (nanowire HEMT, Resonating tunnel diode, etc.)[1,2,3]. Unlike the planer heterostructure where epitaxial III-N layers are grown on foreign substrates (e.g. Sapphire), III-Nitride nanowires exhibit much superior crystal quality owing to their very different growth kinetic and physical structure[4,5]. Further, the advantage of NW is that the dislocations which are primarily originated at the interface due to the lattice mismatch with the substrate don't seem to propagate along the c axis rather stay at the interface and eventually, found to be transferred onto the free side wall[6].

Although NWs in general exhibit superior structural properties due to their very large aspect ratio compare to their planer structure[7] their structural properties strongly depend on the growth condition[8,9]. It is now established that NWs grown at higher temperatures exhibit better optical properties [10]. However, their growth process appears to be inefficient as it takes very long time to start the initial growth of NW, which is known as incubation time [11]. The incubation time is the duration that is required to grow the critical nucleus. The growth of critical nucleus is determined by the available free energy, called "Gibbs free energy" of the system comprising Ga and N adatoms and Si surface. As the Gibb's free energy for the nucleation strongly depends on the size of the nucleus with its maximum at critical value ($r_c$) [12] initial size of the NW nucleus must exceed the critical value in order to continue to the growth of NW (typically known as elongation period). Therefore, the initial process primarily is dominated by both thermodynamic and kinetics. There have been a large number of studies reporting the impact of growth temperature the nucleation and structure of GaN NWs grown by various techniques[13,14,15]. However, structural properties of NW are not yet correlated with the optical properties in a comprehensive manner. In the present study we have carried out a comprehensive investigation on evolution and properties of GaN NWs grown by plasma assisted molecular beam epitaxy technique (PA-MBE) on Si(111) substrates.

**Experimental detail**

All the NWs for the present study have been grown on Si(111) substrates by plasma assisted molecular beam epitaxy (PA-MBE) technique with III-V ratio of 0.4 for all cases. 2 inch Si wafers cleaned by standard RCA cleaning followed by last HF (2%) dip were immediately loaded into the load lock. The wafers were subsequently transferred to the preparation chamber where each wafer was baked for 8 hours at 400°C prior to transferring into the growth chamber. In the growth chamber, the Si (111) wafer was subsequently degassed at 860°C for 30 min where appearance of 7x7 surface superstructure monitored in situ by reflection high energy electron diffraction (RHEED), confirmed clean surface. The substrate temperature during growth was varied from 700°C to 800°C while the III-V ratio (=0.4) was kept constant for all the experiments. The temporal evolution of NWs was monitored in-situ via real time RHEED technique. Microstructural morphology of the NWs was studied by field emission scanning electron microscopy (FEGSEM) and size, shape and distribution from SEM images were analyzed by ImageJ software [16]. The diameter of the NWs varied from 30 -100 nm and length was around 1.0-1.5μm depending on the growth time. The estimated NW density was about $10^9/cm^2$. High resolution X–ray diffraction (HRXRD) measurements were carried out in a Rigaku Smart lab diffractometer equipped with a $Cu_{K\alpha}$ source and a 2-bounce Ge (220) hybrid monochromator. In order to further look into the structural changes with their growth temperature we carried out Raman spectroscopy measurements using the 638 nm excitation line of He-Ne Laser and a Horiba 800 spectrometer. Finally, to study the impact of microstructure on the optical properties, photoluminescence measurements were carried out at different temperatures using continuous-wave (He-Cd) laser with excitation at wavelength of 325nm.

**Results and Discussion**

In spite of being myriad of research done in this subject, there exists no unified and well defined understanding that explains growth mechanism of GaN NW at different stages of growth process. There are number of phenomenological models developed by several well-known groups across the globe to explicate their experimental observation under a particular growth condition [8,17,18,19,20,21]. Among them, the one which is mostly referred has been developed by Gorrido et al.[8]. It is evidenced that with regard to the morphology of the samples, the substrate temperature plays crucial role in transforming two dimensional (2-D) compact layer to one dimensional (1-D) nanowires with same III-V ratio. In present study we observe that sample grown at 700°C exhibits 2D-type islands that extend vertically

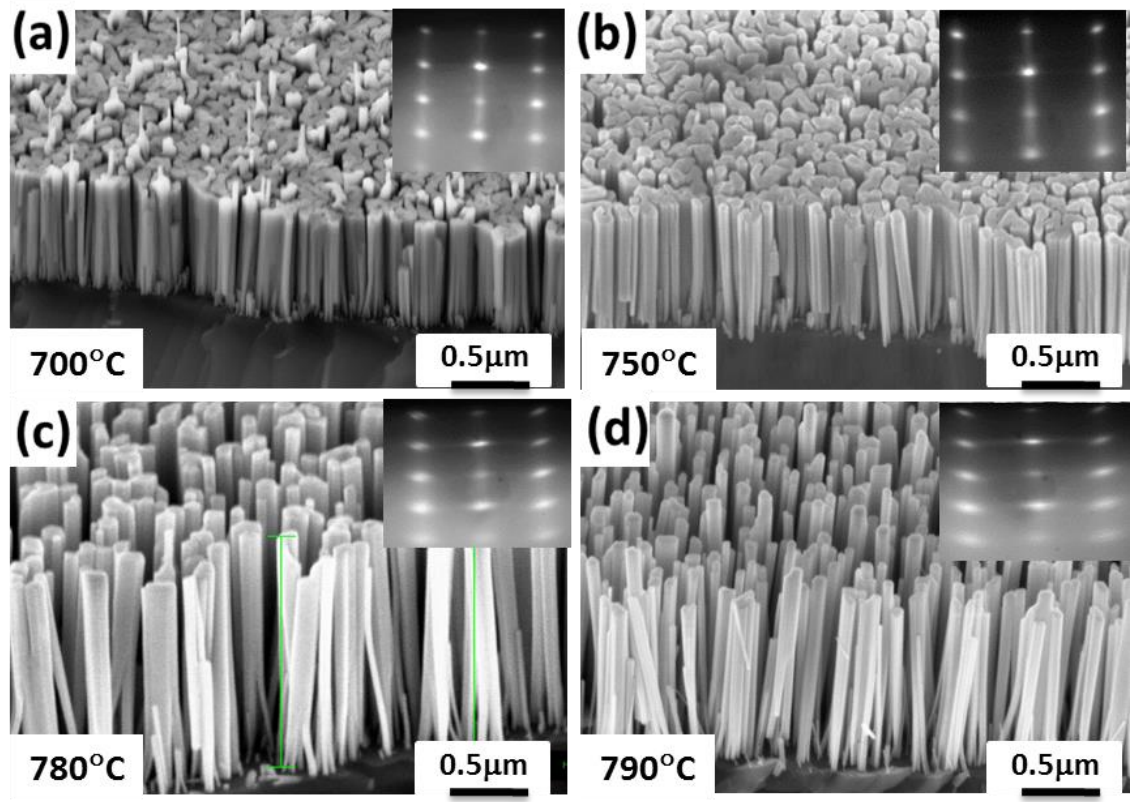

Figure1 (a)-(d). Tilted cross-sectional (bird's eye view) scanning electron micrograph images of samples grown at different temperature illustrating structural evolution of GaN nanowires from a compact layer with columnar structure (700°C grown sample). Insets are the corresponding RHEED images captured in-situ immediately after the growth.

along C-axis with increasing growth duration. However, with increasing growth temperature, 2-D islands evolve into isolated NWs exhibiting excellent structural and optical properties, the details of which will be discussed in the next section. Figure1 compares cross-section view of scanning electron micrograph (SEM) of the samples grown at different substrate temperatures with III-V ration of 0.4.  From the images it is clearly seen that the sample grown at $T_g$= 700°C exhibits 2D-like structure with discontinued long range ordering(viz. columnar grain) inferring that no NW could be grown at this temperature.  As the substrate temperature was increased 2D islands evolve into NWs exhibiting the excellent structural quality of the sample grown at 790°C. Insets in Fig.2 depicts the corresponding reflection high energy electron diffraction (RHEED) image. Interestingly, the sample grown at 700°C and 750°C depict island-like structure with background streaks ascribed to the two dimensional ordering from the compact layer whereas images for the samples grown at 780°C and 790°C show no such ordering, instead reduced lateral dimension[22]. Figures 2(a)–(d) depict the plan-view SEM micrographs of the samples grown at 700°C, 750°C, 780°C and 790°C respectively. It is evidently manifested that the sample grown at 700°C does contain two dimensional islands with almost no isolated texture that could be considered as NW. These isolated grains begin to nucleate when the temperature of the substrate is around 750°C and higher.

 It could further be noticed that with increasing substrate temperature the coverage area is seen to be reduced drastically however their hexagonality is being increased significantly as shown in the insets. This observation agrees well with the thermodynamically stable phase of GaN which possesses the hexagonal symmetry. It is therefore, anticipated that spontaneously formed GaN nanowires would grow along the polar [000$\bar{1}$] direction (C-plane) exhibiting six fold surface symmetry[23,24]. In a further effort to understand the effect of temperature on size, shape and distribution of NW, two basic experiments were carried out where the growth was terminated right after the nucleation. This was carefully monitored using the RHEED technique. Figure 3 presents the tilted and plan view microstructure of samples grown at 700°C (a, b) and 790°C (c, d) respectively. As evident, the sample grown at 700°C appears to have microstructure with multiple grains which are already connected to each other. The plan view image shows 2-D island-like feature (Fig. 3(b)) which continues grow vertically as function of growth time. However, individual NW is already observed right after nucleation for the sample grown at 790°C as shown in Figure 3(c,d). As mentioned earlier, nucleation was monitored using RHEED technique which shows appearance of diffraction images immediately after nucleation, manifesting distinct character that correlates the growth temperatures (in-sets of Fig 3(a) and (c)). The surface of sample grown at 700°C appears to be rough characterized by spotty RHEED image whereas the image from the sample grown at 790°C attributes the feature with significantly reduced lateral dimension [22]. The present results therefore, confirm that the shape and distribution of the NWs are primarily determined by their nuclei geometry.

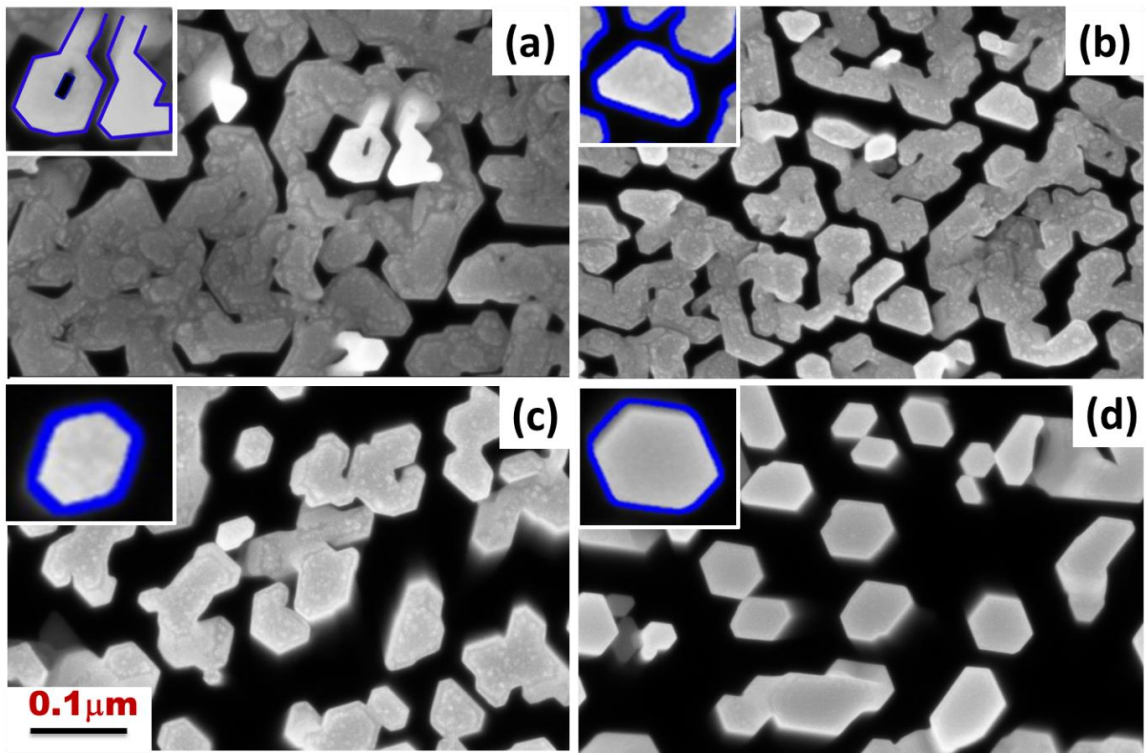

Figure 2. Plan view SEM images for the samples grown at different temperature. Insets represent the magnified images of a single grain/NW depicting the degree of circularity.

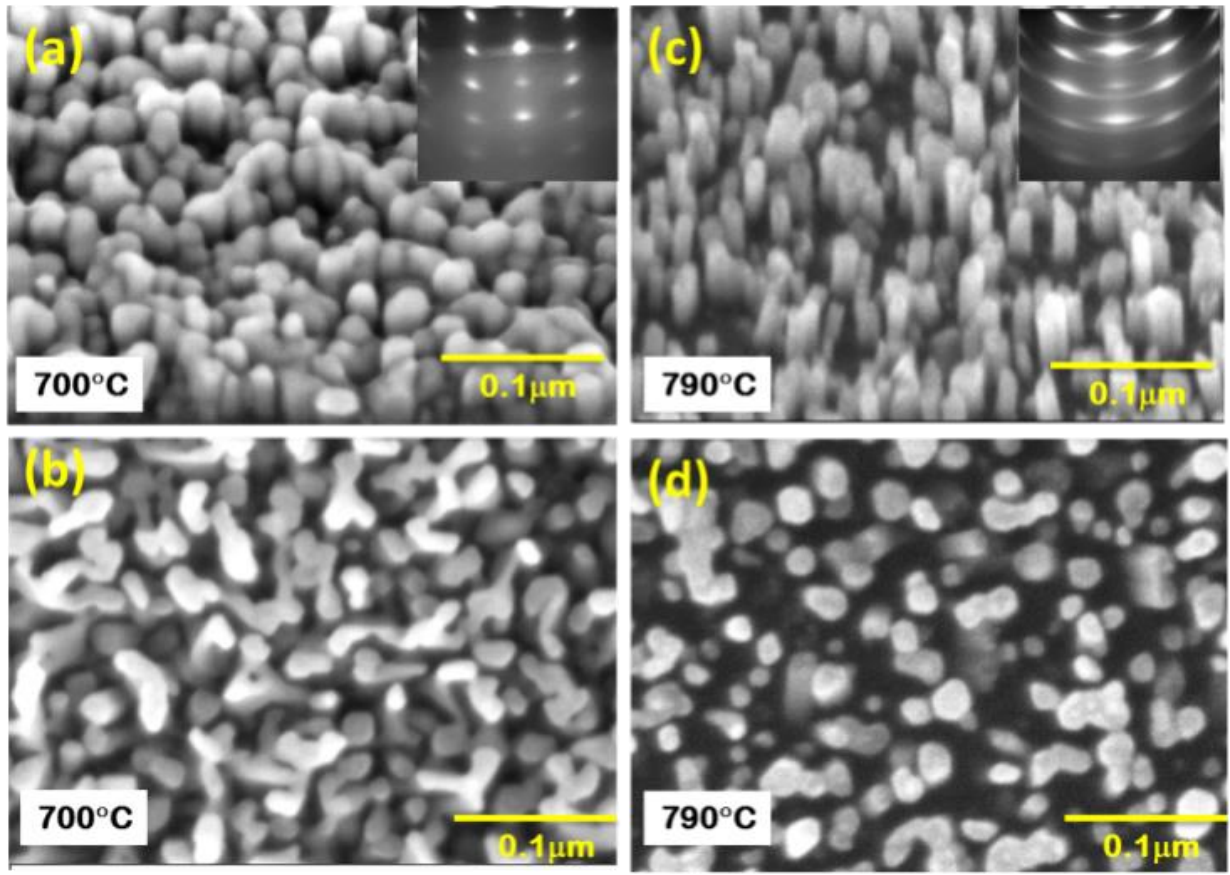

Figure 3: Size, shape and distribution of the GaN NWs soon after nucleated (a) bird view (b) plan view of the sample grown at 700C while (c) (bird view) and (d) (plan view) are the same that grown at 790C.

Another useful parameter that would determine the quality of GaN NW is the circularity (C) or two dimensional isoperimetric quotient [25], which for an object exhibiting arbitrary cross-sectional shape is defined by

$$C = \frac{4\pi A}{p^2},$$

where A is the area and P is the perimeter of the cross-section. The detail analysis of circularity of regular n-gons could be found elsewhere [26]. Figure 4 compares the area coverage (filling factor) and circularity of the present samples as a function of growth temperature. As seen, the NW with best circularity exhibit lowest area coverage, which agrees well with the previously reported results [26].

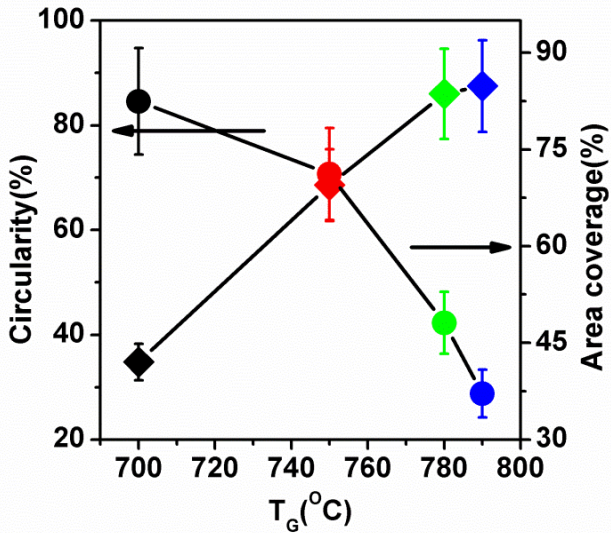

Figure 4. Circularity and area coverage as function of growth temperatures (TG)

Although, most of the growth models available in the literatures work quite well in their respective domains; however, there has not been a single most convincing model that predicts the growth, shape, size and distribution of the critical nucleus that ultimately determines the properties of NW. Earlier, it was believed that Ga droplets onto the substrate surface act as a metal catalyst (like Au, Ti in VLS growth mechanism) for NW growth in PA-MBE system [27]. However in the later year it was investigated and proved that there was no Ga droplet present at the surface or even at the top of the NW even at N rich condition[28].In another work, it was reported that the sticking coefficient of Ga atom on C(0001)plane (tip of the NW) is much higher than the m(1100)-plane(sidewall of NW) at very high temperature in the $N_2$ ambient condition[29] and hence results anisotropic growth which ultimately leads to growth of NW.Further, based on total-energy density functional theory calculations it was estimated that at N-rich surface the Ga ad atom migration barrier is four times higher than the Ga rich condition[30] and therefore is kinetically limited. This was believed to be the primary driving force, inducing the growth of three dimensional nuclei which eventually grow as NW at sufficiently high temperature. This has been mostly accepted and verified mechanism for GaN NW growth. Once the critical nucleus is formed, next Ga adatoms coming onto the surface must diffuse onto the top of the NW from the side wall to minimize the chemical potential [7]. It is necessary that Ga adatoms must have sufficiently large kinetic energy which can only be provided by increasing substrate temperature. However, on contrary, the growth of 2D planer structure is primarily determined by thermodynamics where energetics such as surface free energy, interfacial energy stem from lattice mismatch ultimately decide the crystal quality of the overgrown layer on any foreign substrate.

Although the scope of the present paper is not to dwell on the growth mechanism extensively, nevertheless, we show exclusively that unlike III-V ratio which was believed to be one of the primary parameters, substrate temperature ultimately plays the deciding role on the growth of GaN NW by PA-MBE technique.

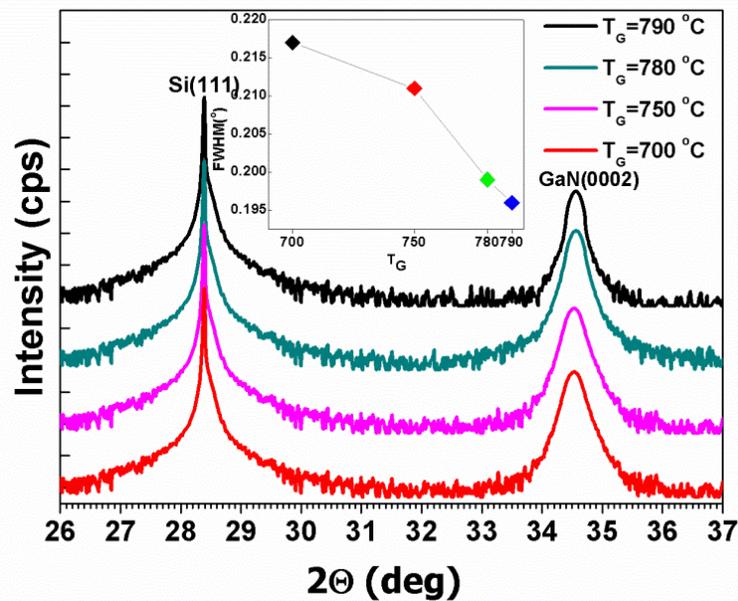

Figure6. The $\Phi$ scans profiles for GaN nanowires at $(10\bar{1}1)$ diffraction peaks

Further, the crystal quality and orientations of the nanowires grown at different temperature are studied from the high resolution x-ray diffraction (HRXRD) scan profiles. Figure 5 shows the wide $\omega - 2\theta$ scan of the GaN nanowires grown at different substrate temperatures. As seen in the figure, only strong (0002) peak of GaN is observed for all the samples at around $34.56^0$. The absence of additional peak confirmed that the nanowires are highly oriented along the c – axis. The full width half maxima (FWHM) of (0002) reflection decreases with increasing growth temperature as shown in the in-set of figure 5. Since the reciprocal space vector associated to the x-rays primarily moves along the out of plane (c – axis) in the symmetric $\omega - 2\theta$ scan, the decrease in FWHM observed in the present case clearly infer that there is a relative improvement in microstructural crystal quality for the NWs grown at higher substrate temperature. Figure 5 compares the azimuthal scan of $(10\bar{1}1)$ reflections for all the samples grown at different temperature. The uniform intensity resulted from all six different $(10\bar{1}1)$ planes due to the six fold symmetry of hexagonal GaN clearly demonstrates excellent in-plane (ab-plane) azimuthal coherence among ensemble of nanowires grown at particular temperature. This measurement once again confirms the structural perfection of the NWs grown at higher temperature.

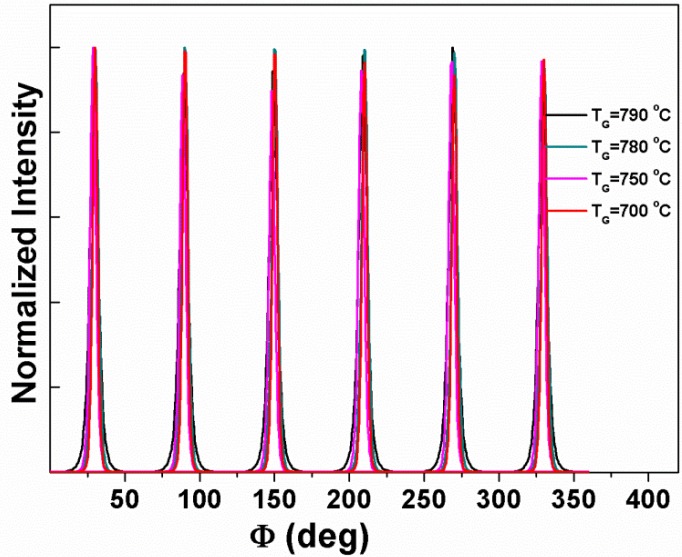

Figure 5. Wide $\omega - 2\theta$ scans profiles for GaN nanowires gown at different temperatures. Inset shows the full width half maxima with function of growth temperature

In order to further look into the structural development with their growth temperature Raman spectroscopy was carried out using the 638 nm excitation line of He-Ne Laser and a Horiba 800 spectrometer. GaN crystal exhibits hexagonal wurtzite structure with four unit cell and belongs to the $C_{6v}$ symmetry group. Group theory predicts two $A_1$, two $E_1$, two $E_2$, and two $B_1$, modes. One $A_1$ and one $E_1$ are acoustic vibrations[31]. The $E_2$ modes ($E_2^{high}$ and $E_2^{low}$) are Raman active, the $A_1$ and the $E_1$ modes are both Raman and Infrared active at the Γ-point of the Brillouin zone of hexagonal GaN, and the $B_1$ mode is silent. $A_1$ and $E_1$ are polar modes and hence they split into two components i.e., $A_1$(LO), $A_1$(TO), $E_1$(LO) and $E_1$(TO). Fig. 7 presents the Raman spectra $E_2^{high}$ phonon peak (only observed mode for GaN NW) of the GaN NWs (length~1μm) and GaN film (thickness: 500nm) grown on Si (111) by PAMBE along the $z(xx)\bar{z}$ scattering geometry. The GaN film grown by PA-MBE exhibit $E_2^{high}$ mode at 566.67. The $E_2^{high}$ phonon peak for the samples grown at 750 to 790 °C is centered at 567.2 ±0.1 cm$^{-1}$, which is consistent with strain free values reported in the literatures[32,33]. We can therefore, infer that all these NWs grown on Si(111) for the present study are free of strain. The $E_2^{high}$ phonon frequency for the NW grown at 700 °C is 566.1 cm$^{-1}$, inferring that the sample grown at lower temperature exhibits characteristic resembling the planer layer. The inset in Fig.7 shows NW $E_2^{high}$ FWHM vs. $T_G$ variation. The $E_2^{high}$ peak FWHM reduces from 5.25 cm$^{-1}$ for $T_G$= 700 °C to 2.65 cm$^{-1}$ for $T_G$= 790 $^o$C thus showing significant improvement in quality of NWs with increase in growth temperature.

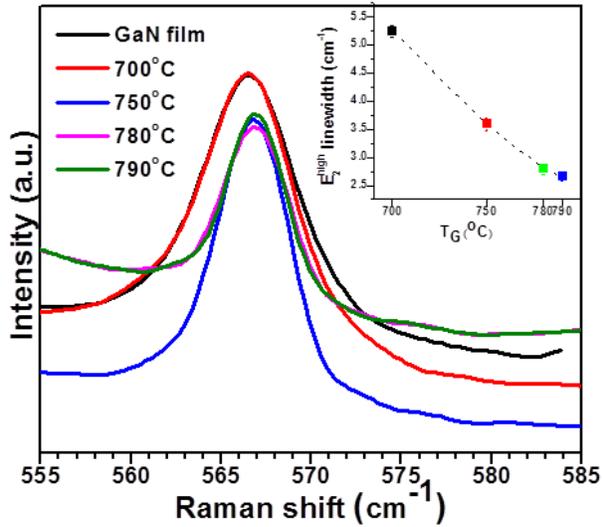

Figure 7. Raman spectra of $E_2^{high}$ peak of NWs and GaN film (Inset: FWHM of $E_2^{high}$ of NWs).

Finally, temperature dependent photoluminescence spectroscopy was carried out on all the samples using He-Cd laser source of wavelength 325nm with excitation energy of 1.5W/cm$^2$. Figure 8 compares the normalized PL spectra measured at room temperature for all the samples exhibiting near band edge (NBE) emission at 3.41eV. As observed, there is significant reduction of PL intensity for the samples grown at lower temperature (shown in the inset) despite all the samples grown for same duration of time, inferring that the microstructure of the NWs exhibits significant influence on their luminescence characteristics. Another important feature that could be manifested in this measurement is the full width half maximum and it was found to reduce with increasing growth temperature shown in the inset of Fig. 8. This once again furnishes the legitimate evidence on the role of microstructure on optical properties.

Finally, the most interesting and important features that could be elucidated in this study, is the excitonic behavior of GaN NWs ensemble measured at lower temperature (10K). Figure 9(a) compares the PL spectra of two samples grown at 700°C and 790°C and a planer GaN layer grown on GaN/Sapphire template by PA-MBE technique. All three spectra exhibit a sharp peak centered at 3.473eV which is ascribed to the recombination of A excitons bound to neutral donors [$(D_0,X_A)$] [27]. The shoulder at 3.48eV (shown in the inset) is usually attributed to the recombination of B-excitons bound to neutral donors$(D_0,X_B)$[34]. Notably, the peak at 3.45 eV is observed only for the sample grown at 790°C and epi-GaN layer however is missing for the sample grown at 700C. There have been several studies reporting various origin of the 3.45eV peak in the literatures. For example, Corfdir et al. have attributed the luminescence at 3.45 eV to two-electron-satellite (TES) excitonic recombination on a near-surface donor [35]. Furtmayret al [36] and Brandt et al. [37] have ascribed this 3.45 eV line to the excitonic recombination on near surface point defects. However, later investigation by D. Sam-Giao et al. using polarization-resolved luminescence and magneto-luminescence experiments contradicted this

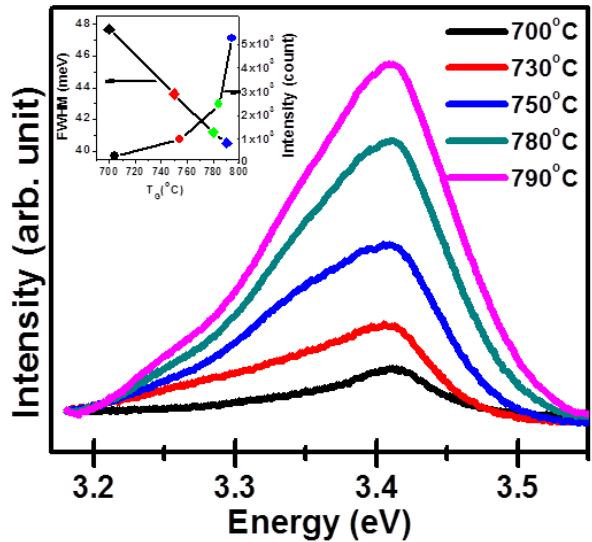

Figure 8. Near Band Edge (NBE) photoluminescence spectra measured at room temperature using 325 nm He-Cd laser with power density of 1.5W/cm2 for the interpretation [38]. Further very recently, Auzelle et al. reported that the 3.45 eV band is actually resulted from the excitonic recombination on a prismatic inversion domain boundaries (pIDB) [39]. First-principles calculations based on many-body perturbation theory combined with polarization-resolved experiments carried out by Pu Huang et al. provides credible evidence that the optical emission from bound excitons localized around the surface microwire to crystal-field split-off hole (CH) band is primarily attributed to the 3.45 eV line with E∥c polarization [40]. Another significant outcome of the low temperature PL measurement is the intensity ratio[I($D_0$,X)/I(I1,X)/]between the transitions corresponding to excitons bound to "neutral donors ($D_0,X$)" and the "I1 basal plane stacking faults"[41,42] for the samples grown at 700°C and 790°C. The values are estimated to be 0.76 and 4.4 for the 700°C and 790°C grown samples respectively, implying that higher temperature is absolutely necessary for growing high quality NWs. Weak bands observed at 3.33eV and 3.36eV manifested only for 790°C grown sample correspond to the LO-phonon replica of the of the (I1,X) and ($D^0X$), respectively. Finally, the $Y_7$ line appeared at 3.21 eV which has been attributed to the emission owing to the recombination of excitons trapped at the boundaries between coalesced nanowires [36,43]. Based on our present measurement and observation we conclude that the sample grown at 700°C behaves like planer structure with large number of extended defects such as stacking faults.

Above all, the most remarkable virtue of GaN NWs compare to the planer structure is their outstanding structural quality which has been witnessed all along in the present investigation. Low temperature PL measurements at the energy range centered at yellow band (broad yellow-band in the visible range`1.8-2.6eV) could provide another insight information about the defect in NWs. Figure 9(b) compares the PL response centered at yellow band for the samples with different microstructures. Interestingly the spectrum belonging to the low temperature (700°C) grown sample resembles the planer structure with fair amount of yellow emission (although the value is much weaker than the planer structure). Just to note here that the curves are shifted along Y-axis for better comparison. Although the origin of the yellow emission in GaN planer layer is still debated in the literatures, the same has been mostly attributed to the structural imperfection induced by several extended defects such as dislocation, grain boundaries and the point defects (e.g. impurities such as C, Si and Ga vacancies $V_{Ga}$) within layer [44,45,46,47]. Therefore, the absence of yellow emission in 790°C grown samples further proves its superior crystal quality. Based on the present results, we can convincingly infer that with same III-V ratio,

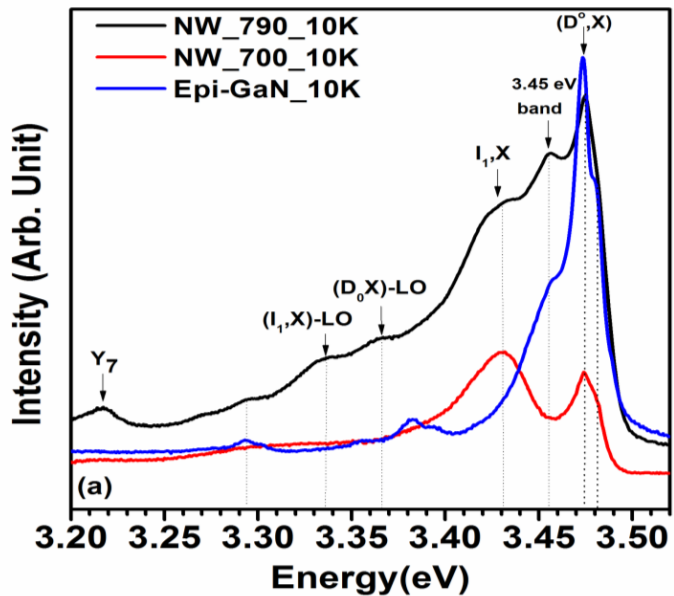
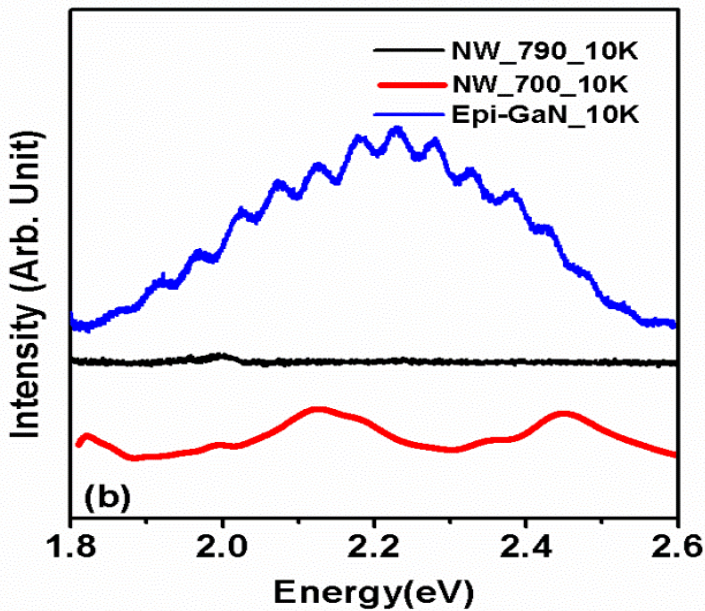

Figure9. (a) Normalized photoluminescence (PL) spectra measured at 10 K for the samples grown at 700°C and 790°C and comparing them with epi-GaN planer layer. (b) Vertically shifted PL spectra of same samples measured in the yellow range (1.8-2.6eV) at 10K.

the substrate temperature should sufficiently high in order to grow GaN NWs with fairly good structural and optical properties.

## Conclusions

In summary, we have clearly demonstrated that the microstructure of GaN NW grown on Si(111) substrates by PA-MBE technique has strong influence on their optical properties. Structural evolution of GaN NWs as shown in this work is primarily determined by the growth temperature. In order to grow the GaN NWs on Si(111) one requires to increase the substrate temperature at least 770°C with III-V ratio is less than one. GaN NWs grown at 790°C exhibits excellent in-plane structural coherence as measured by high resolution XRD and this is evidently reflected in their optical properties. For example, despite having 3 times lower surface area coverage, the absolute intensity of PL spectra of the sample grown at 790°C is one order of magnitude higher than that grown at 700°C. This is further being evidenced in the low temperature PL emission where excitons bound to the I1 basal plane stacking faults(I1,X) dominates over the donor bound excitons($D_0$,X) for the sample grown at 700°C. Whereas no yellow emission owing to such extended defect is visible for the sample grown at 790°C.

## Acknowledgements

All the authors would like to thank the Department Electronic and Information Technology (DEITY), Govt. of India (project code: 11DIT005) for financial support to carry out this work.